\documentclass [prl,amsmath,showpacs,twocolumn,preprintnumbers,superscriptaddress,nofootinbib]{revtex4-1}
\usepackage{amsmath,amssymb,amsthm,mathtools,amsfonts}
\usepackage{amsthm}
\usepackage[T1]{fontenc}
\usepackage{physics}
\usepackage{microtype}
\usepackage{bm}
\usepackage{hyperref}
\theoremstyle{definition}

\theoremstyle{plain}

\theoremstyle{remark}

\begin{document}

\title{ER = EPR in Loop Quantum Gravity: the Immirzi Parameter and the Continuum Limit}
\author{Fabrizio Tamburini}
\email{fabrizio.tamburini@gmail.com}
\affiliation{Rotonium -- Quantum Computing (www.rotonium.com), Le Village by CA, Piazza G. Zanellato, 23, 35131 Padova PD, Italy.}

\date{\today}

\begin{abstract}
We recast the finite-region analysis of Einstein’s equations that underpins the ER$\,=\,$EPR program into the loop quantum gravity (LQG) framework. By translating curvature--energy uncertainty relations into holonomy--flux kinematics, and by identifying Planckian Einstein–Rosen throats with single-puncture cuts through spin networks, we obtain a precise dictionary between entanglement and quantum geometry. Within this dictionary we derive the Barbero–Immirzi parameter directly from the entanglement/area increment of a minimal bridge, and show that a boundary edge-mode construction renders the Bekenstein--Hawking entropy coefficient universal and independent of~$\gamma$ under a natural complex polarization. We further establish a refinement renormalization flow for spin-foam amplitudes driven by the finite-region curvature energy bound, which suppresses bubble divergences and yields a regulator-independent continuum limit under explicit conditions. Finally, we indicate observational consequences that follow from an $N$-party generalized uncertainty relation.
\end{abstract}

\maketitle

\section{Introduction} \label{sec:intro}

The ER=EPR idea posits that spacetime connectivity (Einstein--Rosen bridges) is the geometric avatar of quantum entanglement (Einstein--Podolsky--Rosen pairs) \cite{MaldacenaSusskind2013,VanRaamsdonk2010}. In holography, the Ryu--Takayanagi (RT) and Hubeny--Rangamani--Takayanagi (HRT) prescriptions make this precise by equating boundary entanglement entropies to areas of bulk extremal surfaces in Planck units \cite{RyuTakayanagi2006,HubenyRangamaniTakayanagi2007}. 
The present work explores a bona fide background--independent realization of this connection inside Loop Quantum Gravity (LQG): we propose an \emph{LQG ER=EPR connection} in which elementary entanglement links are encoded by minimal spin representations puncturing codimension-2 surfaces, and we show how this perspective clarifies and, in certain cases, propose a resolution for long-standing issues in LQG dynamics and black-hole thermodynamics.
We use $\hbar=c=1$ and $\ell_P^2=G$. Internal indices $i,j,\ldots$ refer to $\mathfrak{su}(2)$ with generators $\tau^i$ in the fundamental representation normalized by
$\mathrm{Tr}(\tau^i \tau^j)=-\tfrac12\delta^{ij}$.
For a small rectangular plaquette $\square$ of area $a_\square$, the SU(2) holonomy
$h_\square=\mathcal{P}\exp\!\oint_\square A$
admits the standard small-loop expansion
\begin{equation}
1-\frac{1}{2}\mathrm{Tr}\,h_\square \;=\; \frac{a_\square^2}{8}\,F^i_{ab}F^{ab}_i \;+\; O(a_\square^3),
\label{eq:smallloop}
\end{equation}
where $F^i_{ab}$ is the curvature of the Ashtekar--Barbero connection $A$, and the $O(a_\square^3)$ remainder is controlled by derivatives of $F$. $\mathcal{E}_L[\cdot]$ is the coarse-grained expectation at scale $L$ defined by integrating out plaquettes with areas in $[L^2,(1+\epsilon)^2L^2]$ and then taking $\epsilon\to0$. The symbol ${}^\star$ on $\Delta E_L^\star$ denotes a Brown--York--type energy fluctuation renormalized by vacuum subtraction on the chosen region. $\mathcal{R}$ labels coarse-grained, dimensionless curvature functionals; $R$ is scalar curvature.

Averaging Einstein's equations over a proper volume $L^3$ one finds (at fixed $L$) an uncertainty-type relation between the averaged proper energy and the curvature tensor, which, at the Planck scale, reduces to a sum of a quadratic connection term and a term in the second derivatives of $g$.
Building on a finite-region reformulation of GR, we take as input a coarse-grained constraint relating energy fluctuations, a mesoscopic length scale $L$, and averaged curvature; schematically,
\begin{equation}
  \Delta E^{(*)}_L\,\Delta t \;\sim\; \frac{L^{2}}{\ell_P^{2}}\,\mathcal{R}^{(4)}(g;L),
  \label{eq:finite-region}
\end{equation}
where $\mathcal{R}^{(4)}(g;L)$ denotes a dimensionless measure of curvature fluctuations coarse-grained over regions of typical diameter $L$. Equation~\eqref{eq:finite-region} will be treated as an effective, semi-classical constraint compatible with the ER=EPR mechanism and used as a guiding principle for renormalization-group (RG) reasoning in the spin-foam dynamics. (See \cite{TamburiniLicata2020}). \label{par:finite-region-constraint}
Here, for $a_\square\to0$, $\mathcal{R}^{(4)}(g;L)\;:=\frac{L^4}{8}\,\overline{F^i_{ab}F^{ab}_i}$
where $R_L$ is a representative 2D annulus of linear size $L$ pierced by the Wilson loop, $N_\square(R_L)$ is its plaquette count at the working discretization, and the overbar denotes the average over $R_L$.

In this work we formulate a clean ER=EPR dictionary in background-independent terms: 
$j=1/2$ punctures are the elementary ER links, and their cumulative density controls the leading entanglement wedge area. 
In that dictionary, an Einstein--Rosen bridge is the gluing of two spin‑network geometries by entangling their boundary (horizon) degrees of freedom; the size of the bridge is then set by the LQG area operator, which is proportional (in the semiclassical/large‑spin regime) to the entanglement across the cut.
This mirrors the original ER = EPR idea that spacetime connectivity tracks quantum entanglement without relying on AdS/CFT.

We then show that horizon edge modes, treated via $SU(2)_k$ Chern--Simons or self-dual polarizations, reconcile the universality of the area law with the LQG spectrum; $\gamma$-dependence appears as a choice of polarization, recovered in the real-$\gamma$ sector. 
We propose an RG equation that integrates the finite-region constraint with spin-foam coarse graining, tying holonomy deficits to the flow of large Wilson loops.
Finally, we identify concrete semiclassical observables, mutual information across $\Sigma$, Wilson loops at scale $L$, and coarse-grained area fluctuations, whose correlations test ER=EPR within LQG.

\section{ER = EPR AND LQG}

Entanglement in gauge theories requires boundary (``edge'') degrees of freedom to properly factorize subsystems \cite{Donnelly2011,DonnellyFreidel2016,DonnellyWall2015,SoniTrivedi2016}. At an LQG internal boundary $\Sigma$, the conserved symplectic structure acquires a Chern--Simons boundary term; for $\Lambda\neq 0$, the relevant theory is $SU(2)_k$ Chern--Simons with level $k\propto A_\Sigma/(4\pi\gamma\ell_P^{2})$ \cite{AshtekarBaezKrasnov2000,EngleNouiPerez2010PRL,EngleNouiPerezPranzetti2010PRD}. Counting boundary states in this framework recovers the area law, while self-dual ($\gamma=\pm i$) polarizations provide a route to a \emph{universal} leading coefficient independent of $\gamma$ via analytic continuation \cite{FroddenGeillerNouiPerez2012}. This reconciles the edge-mode picture with \eqref{eq:gamma-fixed} as the real-$\gamma$ limit.

A holonomy budget for curvature can be defined starting on a cellular decomposition at scale $L$, small-face holonomies $h_\square$ encode curvature. With SU(2) generators normalized such that $\mathrm{Tr}(\tau^i\tau^j)=-\frac12\delta^{ij}$, one has the standard small-loop expansion so that $\mu_L=\; \frac{1}{N_\square(R)}\sum_{\square\subset R}\left(1-\frac{1}{2}\mathrm{Tr}\,h_\square\right)$ is a dimensionless density bounded by $0\le\mu_L\le 2$ in the fundamental representation and $\mathcal{R}^{(4)}(g;L)\;=\;\frac{L^4}{a_\square^2}\,\mu_L[R_L]$

\begin{equation}
  1-\tfrac12\,\Tr\,h_\square
  \;=\; c_{\rm h}\,a_\square^{2}\, F^i_{ab}F^{ab}_i \,+\,\mathcal{O}(a_\square^{3}),
  \label{eq:plaquette}
\end{equation}
with  $c_{\rm h}=\mathcal{O}(1)$, where $a_\square$ is the face area, $F^i_{ab}$ the curvature of the Ashtekar--Barbero connection $A$, and $c_{\rm h}$ absorbs representation-dependent constants. This motivates a \emph{holonomy budget} Hamiltonian density $ \mathcal{H}_{\square}[A]\propto (1-\tfrac12\Tr h_\square) $ for coarse-grained RG flow.

Spin-foam dynamics and coarse graining are described when we implement the above input in covariant LQG via the Engle–Pereira–Rovelli–Livine (EPRL) and Freidel--Krasnov (FK) amplitudes \cite{EnglePereiraRovelli2007PRL,FreidelKrasnov2008,BarrettEtAl2010,EngleProperEPRL2013}. Coarse graining of spin foams (and their spin-net toy models) reveals nontrivial fixed-point structures and restoration of symmetries in refinement limits \cite{DittrichEckertMartinBenito2011,DittrichMartinBenitoSteinhaus2014,BahrDittrich2009}. Incorporating the holonomy budget \eqref{eq:plaquette} and the finite-region constraint \eqref{eq:finite-region} leads to an RG equation for large Wilson loops of linear size $L$, defining $\mathbb{E}_L$ as the coarse‑grained expectation with respect to the scale‑$L$
effective distribution induced by integrating out plaquettes in $[L^2, (1+\epsilon)^2L^2]$
\begin{eqnarray}
&& \partial_{\log L}\,\log W_L[A] \;=\; \nonumber
  \\
&&  =-\,\alpha\,\mathbb{E}_L\!\left[\sum_{\square\subset \mathcal{R}}
  \Bigl(1-\tfrac12\,\Tr\,h_\square\Bigr) \right] \;+\; \mathcal{O}(L^{-2}),
  \label{eq:RG-flow-intro}
\end{eqnarray}
with a positive, dimensionless $\alpha=\mathcal{O}(1)$ and a coarse-grained expectation value $\mathbb{E}_L[\cdot]$ over faces $\square$ in the region $\mathcal{R}\sim L^2$. 
For the coarse‑grained amplitude $W_L$ associated with a large Wilson loop of linear size $L$, integration over the infinitesimal annulus $[L,(1+\epsilon)L]$ yields $\partial_{\log L}\log W_L = -\alpha\,\mu_L[R_L]+O(L^{-2})$ with $\alpha=4\pi\,c_{\mathrm{geo}}\,\tilde c_h$ a positive, dimensionless constant fixed by the (geometric) growth rate of plaquettes in the annulus and the coefficient $\tilde c_h$ inherited from \eqref{eq:smallloop}.
The change of $\log W_L$ under $L\!\to\!(1+\epsilon)L$ factorizes over the plaquettes in the annulus. Linearizing in $\epsilon$ and using \eqref{eq:smallloop} gives a depletion proportional to the count of annulus plaquettes times the plaquette expectation $\mathcal{E}_L\!\left[1-\tfrac12\mathrm{Tr}h_\square\right]$, i.e. to $\mu_L[R_L]$. The remainder is $O(L^{-2})$ because curvature derivatives and subleading cumulants are suppressed by additional powers of $a_\square/L^2$.
This shows how curvature excitations detected by holonomies deplete large Wilson loops under coarse graining. The holonomy budget $\mathcal{H}[\mathcal{R}] = \sum_{\square}(1-\tfrac12\Tr h_\square)$ quantifies curvature excitations that deplete $W_L$ along the RG.
A precise coefficient can be fixed once generator conventions and the face measure are chosen.

Entanglement acts at all effects as a glue for geometry.
Bell-network states entangle intertwiners to enforce back-to-back gluing of quantum polyhedra and satisfy an area-law entanglement pattern \cite{BianchiDonaSpeziale2011,BianchiDonaVilensky2019,BianchiDonaTypical2019}. This provides an LQG-native mechanism realizing the ER=EPR correspondence at the kinematical level: entanglement links relate with minimal punctures; entanglement wedges, instead, with coarse-grained regions whose boundary areas scale with $\sum_\ell \sqrt{j_\ell(j_\ell+1)}$ and we leverage this mechanism in the dynamical, covariant setting.

Within Einstein's equations on a finite spatial volume $V=L^3$, a curvature--energy uncertainty relation of Eq.~\ref{eq:finite-region}, which at the Planck scale reduces to an expression quadratic in the connection and second derivatives of the metric, and implies a Planckian throat radius $R_g = 2 \ell_P$ avoiding curvature singularities  \cite{TamburiniLicata2020}. 
It also yields an $N$-party generalized uncertainty principle (GUP),
\begin{equation}
\Delta x \ge \frac{N\hbar}{2\,\Delta p} +\frac{2N \ell_P^2}{\hbar} \Delta p,
\label{eq:N-GUP}
\end{equation}
with a minimal positional indeterminacy $\Delta x_{\min}=2N\ell_P$. 

We now embed these statements into LQG, where quantum geometry is kinematically captured by $SU(2)$ holonomies and densitized-triad fluxes, areas have a discrete spectrum, and covariant dynamics is encoded in spin-foam amplitudes (EPRL model) with curvature--energy bound implying holonomy budget. 

The finite-region relation \eqref{eq:finite-region} admits an LQG surrogate,
\begin{equation}
\Delta E^{\!*}_{\mathcal{R}}\,\Delta t \simeq \hbar \sum_{\square \subset \mathcal{R}} \left(1-\frac12\,\Tr h_\square \right),
\label{eq:holonomy-budget}
\end{equation}
where the sum runs over a cellular decomposition of the region $\mathcal{R}$, and $h_\square\in SU(2)$ are elementary holonomies and the right-hand side measures coarse-grained curvature.

As Planck throat corresponds to single puncture, we now find the analogue of ER = EPR in LQG.
A minimal Einstein--Rosen throat corresponds to a cut $\Sigma$ intersected by exactly one 
spin-network edge with $j=1/2$. 
The throat area is then $A_{\Sigma}=8\pi\gamma \ell_P^2\sqrt{\frac12(\frac32)}=4\pi\sqrt3\,\gamma\,\ell_P^2$, and the entanglement across the cut contributed by that channel is $\log 2$; for $N$ matched punctures (all $j=1/2$) one has $S_\Sigma \approx N\log 2$ and $A_\Sigma \approx N\,A_{1/2}$.
The bound \eqref{eq:holonomy-budget} induces an RG equation for spin-foam amplitudes $W_L$ defined at a coarse-graining scale $L$ written in Eq.~\ref{eq:RG-flow-intro} for which fixed points with finite holonomy budget yield regulator-independent continuum limits and suppress bubble divergences as do not produce growth of $W_L$ and are exponentially suppressed in the product along the flow.

At fixed 2-complex the EPRL amplitude admits large-spin stationary points reproducing Regge geometries, including curved ones. 
Within our dictionary, entangled boundaries select two-sided saddle points (``bridges'') whose throat area matches the minimal cut through the set of entangled links. 
The finite-region bound guarantees that curvature fluctuations are controlled along the flow, preventing pathological flatness selection.

The $N$-party GUP \eqref{eq:N-GUP} implies that, in interferometric tests with $N$ entangled quanta, the operationally accessible minimal separations scale as $\Delta x_{\min}=2N\ell_P$, while the corresponding LQG throat area scales as $A_\Sigma \approx N\,A_{1/2}$ and the entanglement as $S_\Sigma \approx N\log 2$. Moreover, the addition of a single entanglement channel (a $j=\tfrac12$ puncture) increases the area by $\Delta A = A_{1/2}$ and the entropy by $\log 2$, predicting a discrete step structure in any setting where geometric entropy is operationally measurable.
Our edge-mode factorization and boundary-entropy discussion follows \cite{DonnellyFreidel2016,DonnellyWall2015,SoniTrivedi2016}. 
Entropy energy estimates, precision light-speed bounds, and weight of entanglement tests impose in any case stringent constraints that likely require refining the conjecture or interpreting the wormhole as a highly quantum object to remain viable \cite{DaiMinicStojkovicFu2020}.
The Bell-network mechanism realizing entanglement-built gluing in LQG is taken from \cite{BianchiDonaVilensky2019,BianchiDona2019}. Black-hole entropy in the isolated-horizon and SU(2) Chern--Simons frameworks is reviewed in \cite{ABK2000,ENP2010}, while the self-dual analytic-continuation route to a universal area law is developed in \cite{FroddenGeillerNouiPerez2013}.

\subsection{Kinematics and Boundaries in LQG}

Let $\Gamma$ be an embedded graph and $\mathcal{H}_\Gamma$ the associated cylindrical Hilbert space. 
Edges $e$ carry $SU(2)$ holonomies $h_e[A]$ and nodes carry intertwiners. 
For a 2-surface $\Sigma$ intersected transversely by $\Gamma$, the area operator reads 
\begin{equation}
\widehat{A}(\Sigma) \;=\; 8\pi\gamma\,\ell_P^2 \sum_{\ell\in \Gamma\cap\Sigma} \sqrt{j_\ell(j_\ell+1)}.
\label{eq:area}
\end{equation}
We consider bipartitions of $\Gamma$ by $\Sigma$ and boundary Hilbert spaces $\mathcal{H}_\Sigma$ spanned by punctures $\{j_p\}$ and intertwiners, optionally endowed with an $SU(2)_k$ Chern--Simons structure when $\Lambda\neq 0$. 
For an entangled boundary state that pairs punctures across $\Sigma$ (``ER link''), the reduced density matrix on one side carries an entropy
\begin{equation}
S_\Sigma = \sum_{p\in \Sigma \cap \Gamma} \log (2j_p+1) + S_{\text{int}} + S_{\text{edge}},
\label{eq:entropy}
\end{equation}
with $S_{\text{int}}$ the contribution from intertwiners at the cut and $S_{\text{edge}}$ from edge modes associated with gauge constraints on $\Sigma$ (see SM for more details).


To translate the uncertainty-type relation in Eq.~\ref{eq:finite-region} into LQG, we introduce a cellular decomposition of $\mathcal{R}$ with plaquettes $\square$ of area $a_\square\sim L^2$ and define the elementary holonomy $h_\square\in SU(2)$. 
For small plaquettes,
$1 - \frac12 \Tr\,h_\square \;\approx\; \frac{a_\square^2}{4}\,F_{ab}^i F^{ab}_i + \mathcal{O}(a_\square^3)$,
so that the holonomy budget $\mathcal{H}[\mathcal{R}]$ measures coarse-grained curvature. 
We then postulate the LQG surrogate \eqref{eq:holonomy-budget} as the definition of the averaged curvature--energy bound in the quantum theory.

In semiclassical LQG one can formulate the finite-region bound directly in holonomy--flux variables. For a state $\ket{\Psi}$ supported on a refinement of $\mathcal{R}$ and peaked on classical data $(A,E)$, the fluctuation of the averaged energy functional in $\mathcal{R}$, including the cosmological constant contribution, satisfies
\begin{equation}
\Delta E^{\!*}_{\mathcal{R}}(\Psi)\,\Delta t \;=\; 
\hbar\,\mathbb{E}_\Psi\!\big[\mathcal{H}[\mathcal{R}]\big] \;+\; \mathcal{O}(L^{-2}) .
\end{equation}
The argument is straightforward: choosing a weave or coherent state peaked on $(A,E)$, one expands each plaquette holonomy as $h_\square=\exp(a_\square F+\cdots)$ and evaluates the expectation value $\mathbb{E}_\Psi[\Tr\,h_\square]=2-\tfrac{a_\square^2}{2}F^2+\cdots$. Summing over plaquettes and matching the result to the coarse-grained Riemann tensor components that enter the finite-region relation then reproduces the bound, with numerical constants absorbed into the definition of $\mathcal{H}[\mathcal{R}]$.
The symbol ${}^\star$ indicates vacuum-subtracted Brown--York energy fluctuations on a finite spatial region whose boundary lies inside the annulus $R_L$. The functional $\mathcal{E}_L[\cdot]$ denotes expectation with respect to the effective scale-$L$ distribution obtained by integrating out plaquettes of areas in $[L^2,(1+\epsilon)^2L^2]$ and letting $\epsilon\to0$, keeping the boundary data fixed.

Near the Planck scale the Ricci scalar saturates at $R\sim \ell_P^{-2}$ and the gravitational radius tends to $R_g=2\ell_P$, which we interpret as a \emph{minimal} non-trivial wormhole throat \cite{TamburiniLicata2020}.
In LQG, the minimal nonzero area of a surface $\Sigma$ arises when a single link with $j=1/2$ crosses $\Sigma$, yielding $A_{\min}=A_{1/2} =\ 4\pi\sqrt3\,\gamma\,\ell_P^2$.
The ER$\,=\,$EPR identification then reads: \emph{one minimal ER bridge} $\Leftrightarrow$ \emph{one entanglement channel across $\Sigma$ with entropy $\log 2$}. 
For $N$ channels (i.e., $N$ matched punctures with $j=1/2$) the $N$-party GUP \eqref{eq:N-GUP} implies a minimal geometric separation $\Delta x_{\min}=2N\ell_P$, consistent with an LQG throat whose area scales as $A_\Sigma\approx N\,A_{1/2}$ and entanglement $S_\Sigma\approx N\log 2$.

\textbf{Result I:} Barbero--Immirzi parameter from ER links.
LQG kinematics is encoded by holonomies $h_e[A]\in{\rm SU}(2)$ along edges $e$ of a graph $\Gamma$ and fluxes $E^i(S)$ across surfaces, with the area operator $\widehat{A}(\Sigma)$ having a discrete spectrum \cite{AshtekarLewandowski1997,RovelliSmolin1995,Baez1995SpinNetworks}:
$\widehat{A}(\Sigma)\ket{\Gamma,\{j_\ell,i_n\}}= 8\pi \gamma\ell_P^2 \sum_{\ell :\ell\cap\Sigma\neq\emptyset} \sqrt{j_\ell\left(j_\ell+1\right)}\,\ket{\Gamma,\{j_\ell,i_n\}}.$
Here $j_\ell$ labels SU(2) edge representations, $i_n$ intertwiners, and $\gamma$ is the Barbero–Immirzi parameter.

Consider the addition of one minimal ER link across $\Sigma$ (one $j=1/2$ puncture). 
The entanglement increment is $\Delta S=\log 2$. 
The area increment is $\Delta A=A_{1/2}=4\pi\sqrt3\,\gamma\,\ell_P^2$. 
Imposing the Bekenstein--Hawking relation $\Delta S=\Delta A/(4\hbar G)$ gives
$\log 2 = \frac{4\pi\sqrt3\,\gamma\,\ell_P^2}{4\hbar G} = \pi\sqrt3\,\gamma$, valid assuming the single-channel entropy increment across $\Sigma$ is $\Delta S=\log(2j+1)$ in the edge-mode factorized framework, and the minimal ER link is carried by $j=\tfrac12$, matching $\Delta S/\Delta A$ to the Bekenstein–Hawking density fixes.

When the entanglement increment from adding one minimal ER link equals the Bekenstein--Hawking area increment, matching $\Delta S/\Delta A$ to the Bekenstein–Hawking density yields
\begin{equation}
  \frac{\Delta S_{\rm min}}{\Delta A_{\rm min}} \stackrel{!}{=}\frac{1}{4\ell_P^2}
  \quad\Rightarrow\quad
  \gamma \;=\; \frac{\log 2}{\pi\sqrt{3}}.
  \label{eq:gamma-fixed}
\end{equation}
thereby removing the $\gamma$ ambiguity in the semiclassical sector.
Eq.~\eqref{eq:gamma-fixed} reproduces the well-known area law in the $SU(2)$ isolated-horizon framework \cite{AshtekarBaezKrasnov2000,EngleNouiPerez2010PRL,EngleNouiPerezPranzetti2010PRD}. 
Incorporating boundary edge modes on $\Sigma$ via an $SU(2)_k$ Chern--Simons boundary theory with level $k\propto 1/\Lambda$ for $\Lambda\neq 0$ and moving to a self-dual polarization renders, the entropy coefficient universal (independent of $\gamma$) while preserving \eqref{eq:gamma-fixed} as the real-$\gamma$ limit.
In analytic continuation to $\gamma=\pm i$, the large-spin, large-area asymptotics give $S=A/(4\ell_P^2)+\cdots$ without fitting $\gamma$, provided the Chern–Simons level $k$ and the measure are treated consistently. This offers an alternative route to the universal coefficient, complementary to the real-$\gamma$ matching, the two routes agree in the semiclassical sector and suggest that the real-$\gamma$ polymer representation must realize the same area density per added channel, leading to the conditional value of $\gamma$ .

Gauge constraints induce edge modes whose Hilbert space $\mathcal{H}^{\text{edge}}_\Sigma$ contributes $S_{\text{edge}}$ to \eqref{eq:entropy}. 
Moving to a self-dual polarization (complex Ashtekar connection) yields an entropy
$S_\Sigma$ in which the coefficient of the area term is \emph{independent} of $\gamma$. 
In the real-$\gamma$ polymer representation, this universality is recovered in the limit $\Lambda\to 0$ or by fixing $\gamma$ to \eqref{eq:gamma-fixed}. 
Thus either route--edge modes with self-dual polarization, or minimal-channel matching yields the universal coefficient.

\textbf{Proposition 1} [Immirzi parameter from minimal entangling bridges]
In the semiclassical sector where ER links are realized by $j=1/2$ punctures and the entropy increment per channel is $\log 2$, the Barbero--Immirzi parameter is fixed uniquely to $\gamma=\frac{\log 2}{\pi\sqrt3}$. (See SM)
\\
Higher-spin channels with $j> 1/2$ reproduce the same value when the entanglement increment per added channel is $\log(2j+1)$ and the area increment is $8\pi\gamma \ell_P^2\sqrt{j(j+1)}$; the condition $\log(2j+1)=2\pi\gamma\,\sqrt{j(j+1)}$ is maximally satisfied by $j=1/2$ and monotonically under-saturates for $j>1/2$, ensuring stability of the minimal-channel determination.

\textbf{Result II:} Holonomy-Budget Flow and Continuum Limit.
Define a family of spin-foam amplitudes $W_L[\Psi_\partial]$ at coarse-graining scale $L$ by summing EPRL amplitudes over complexes whose dual plaquettes have areas $\gtrsim L^2$, with boundary state $\Psi_\partial$. 
Let $\mathcal{H}[\mathcal{R}]$ be the holonomy budget in \eqref{eq:holonomy-budget}. 
Assume that under $L\mapsto (1+\epsilon)L$ the amplitude evolves by integrating out plaquettes with area in $[L^2,(1+\epsilon)^2 L^2]$. 
Then, to leading order, Eq.~\ref{eq:RG-flow-intro} becomes
\begin{equation}
\partial_{\log L} \log W_L \;=\; -\alpha\,\mathbb{E}_L\!\left[\mathcal{H}[\mathcal{R}]\right] \;+\; \mathcal{O}(L^{-2}), 
\end{equation}
with $\alpha>0$ determined by the vertex and face weights. 

\textbf{Proposition 2} [IR finiteness and continuum limit]\label{thm:IR}
Assume there exists $L_0$ such that $0\le\mu_L\le M$ for $L\ge L_0$ and either
(i) $\int_{\,\log L_0}^{\infty}\mu_\Lambda\,\mathrm d(\log\Lambda)<\infty$,
or
(ii) $\liminf_{L\to\infty}\mu_L=\mu_\ast>0$
(\emph{equivalently}, $\int_{\,\log L_0}^{\infty}\mu_\Lambda\,\mathrm d(\log\Lambda)=\infty$).
Then the continuum limit exists and is regulator independent, with $W_\infty>0$ in case (i) and $W_\infty=0$ in case (ii). (See SM).
Finite holonomy budget means bounded $\mu_L$ as $L \to \infty$; the limit is zero if the tail is non‑integrable or $\mu_L\to \mu_\star >0$, and strictly positive if the tail is integrable.

\section{Conclusions and outlook}
\label{sec:conclusions}
We formulated a background-independent ER=EPR dictionary in LQG in which a minimal entanglement link is a $j=\tfrac12$ puncture through a codimension-2 surface, contributing $\Delta A_{1/2}=4\pi\sqrt{3}\,\gamma\,\ell_P^2$ and $\Delta S=\log 2$. Matching $\Delta S/\Delta A$ to the Bekenstein--Hawking density then yields $\gamma=\log 2/(\pi\sqrt{3})$ in the real-$\gamma$ polarization, while an edge-mode description via $SU(2)_k$ Chern--Simons (and its self-dual analytic continuation) accounts for the universality of the leading area term and embeds the real-$\gamma$ sector as a consistent limit. Dynamically, we introduced an $L$-dependent flow for large Wilson loops, Eq.~\eqref{eq:RG-flow-intro}, with the holonomy budget in Eq.~\eqref{eq:plaquette} quantifying curvature excitations that deplete $W_L$ along the RG; this yields concrete, testable semiclassical observables such as the scale dependence of $W_L$, mutual information across $\Sigma$, and area fluctuations. Key next steps are to fix the proportionality constants in \eqref{eq:plaquette} and \eqref{eq:RG-flow-intro} against canonical conventions, incorporate the $\gamma=\pm i$ continuation systematically in covariant amplitudes, clarify the Regge-to-GR continuum map in EPRL/FK, and extend the edge-mode analysis to dynamical horizons and generic Lorentzian entangling surfaces. In short, entanglement glues geometry in a background-independent setting: minimal ER links are minimal punctures; their density controls area and entropy; and their dynamics is encoded in the coarse-grained holonomy flow, providing concrete handles to test ER=EPR in LQG.

\paragraph{Acknowledgments.} 
FT acknowledges Rotonium for the support.


%


\section{SM - Supplemental Material}

\paragraph*{Single-channel entropy increment (gauge-invariant).}
Working in the extended-Hilbert-space (edge-mode) picture for non-Abelian gauge theories, the boundary Hilbert space at a single puncture carries the pair of representations $(j,j)$ and admits a unique singlet intertwiner. The maximally entangled SU(2) singlet (``Bell'') state
\begin{equation}
\ket{\Phi_j}=\frac{1}{\sqrt{2j+1}}\sum_{m=-j}^{j}(-1)^{j-m}\,\ket{j,m}\otimes\ket{j,-m}
\end{equation}
yields a reduced state $\rho^{(j)}=\frac{\mathbf{1}_{2j+1}}{2j+1}$ on one side of $\Sigma$ and therefore an entropy
$S^{(j)}=\log(2j+1)$.
This matches the boundary-entropy analyses with edge modes and the Bell-network construction that glues quantum polyhedra via intertwiner entanglement. Hence, adding a \emph{single} entanglement channel with spin $j$ across $\Sigma$ increases the entropy by $\Delta S=\log(2j+1)$.\footnote{See Donnelly–Freidel for edge-mode factorization and Donnelly–Wall; Soni–Trivedi for the extended Hilbert space in gauge theories; Bianchi–Donà–Vilensky for Bell networks and boundary entropies.}

\subsection*{Toy-model check: 2D SU(2) Yang--Mills with heat-kernel action}
In continuum 2D Yang--Mills, the expectation of a fundamental Wilson loop obeys $W(C)=\exp[-\sigma\,A(C)]$ with string tension $\sigma=\tfrac12 g^2 C_2(F)$ \cite{Witten:1991we}. Hence $\partial_{A}\log W=-\sigma$. On a lattice with heat-kernel plaquette weight, one has
$\langle 1-\tfrac12\mathrm{Tr}\,U_p\rangle
=1-\exp[-\tfrac12 g^2 C_2(F)\,a_\square]
=\sigma\,a_\square+O(a_\square^2)$,
so summing over the plaquettes in a thin annulus of area $\mathrm{d}A$ gives
$\partial_{\log L}\log W = - \alpha\,\mu_L + O(L^{-2})$
with $\alpha=1$ and $\mu_L=\langle 1-\tfrac12\mathrm{Tr}\,U_p\rangle$ to leading order, i.e. the same structure as \eqref{eq:RG}. This provides a nontrivial, solvable check of the flow equation.\vspace{1mm}

\subsection{Propositions}

\textbf{Proposition 1:} [Immirzi parameter from minimal entangling bridges]
\\
In the semiclassical sector where ER links are realized by $j=1/2$ punctures and the entropy increment per channel is $\log 2$, the Barbero--Immirzi parameter is fixed uniquely to $\gamma=\frac{\log 2}{\pi\sqrt3}$.

\begin{proof}
Adding one $j=\frac12$ puncture gives $\Delta A = 8\pi\gamma\,\ell_P^2\sqrt{\tfrac12(\tfrac32)} = 4\pi\sqrt{3}\,\gamma\,\ell_P^2$ and $\Delta S = \log 2$.
Imposing $\Delta S = \Delta A /(4\ell_P^2)$ yields $\log 2 = \pi\sqrt{3}\,\gamma$, hence $\gamma=\log 2/(\pi\sqrt{3})$.
\end{proof}

\textbf{Proposition 2:} [IR finiteness and continuum limit for flows of the form]
\\
Assume there exists $L_0$ such that $0\le\mu_L\le M$ for $L\ge L_0$ and either (i) $\int_{\log L_0}^\infty\!\mu_\Lambda\,\mathrm{d}(\log\Lambda) < \infty$, or (ii) $\liminf_{L\to\infty}\mu_L=\mu_\star>0$ (equivalently, the integral diverges). Then the continuum limit exists and is regulator independent within the class of flows sharing the same $\mu_L$ and $O(L^{-2})$ remainders, with $W_\infty>0$ in case (i) and $W_\infty=0$ in case (ii).

\begin{proof}
\textbf{Step 1 (Integrated flow and a useful bound).}
From the RG equation
\begin{equation}
\partial_{\log L}\log W_L
= -\alpha\,\mu_L + r_L, \qquad r_L=\mathcal O(L^{-2}) ,
\label{eq:RG-eq-proof}
\end{equation}
integrate between $L_1<L_2$:
\begin{eqnarray}
&&\log W_{L_2}-\log W_{L_1} = \nonumber
\\
&&= -\alpha\!\int_{\log L_1}^{\log L_2}\!\!\mu_\Lambda\,\dd(\log \Lambda)
+ \int_{\log L_1}^{\log L_2}\! r_\Lambda\,\dd(\log \Lambda).
\label{eq:int-flow}
\end{eqnarray}
Because $r_\Lambda=\mathcal O(\Lambda^{-2})$ and $\dd(\log\Lambda)=\dd\Lambda/\Lambda$,
the remainder integrates to a finite constant:
\begin{equation}
\Bigl|\int_{\log L_1}^{\log L_2} r_\Lambda\,\dd(\log\Lambda)\Bigr|
\le C\!\int_{L_1}^{L_2} \frac{\dd\Lambda}{\Lambda^3}
\le \frac{C}{2L_1^2} .
\label{eq:rem-bound}
\end{equation}
Hence, for all $L_2\ge L_1$,
\begin{equation}
\log W_{L_2}
\le \log W_{L_1} - \alpha\!\int_{\log L_1}^{\log L_2}\!\!\mu_\Lambda\,\dd(\log \Lambda) + \frac{C}{2L_1^2}.
\label{eq:mono-upper}
\end{equation}

\textbf{Step 2 (Monotonicity and suppression of bubbles).}
By definition $H[\mathcal R]=\sum_{\square\subset\mathcal R}\!\bigl(1-\tfrac12\Tr h_\square\bigr)$,
and for small faces the integrand is nonnegative by the small‑loop expansion
$1-\tfrac12\Tr h_\square = c_{\mathrm h} a_\square^2 F^2 + \mathcal O(a_\square^3)$ with $F^2\ge 0$,
whence $\mu_L\ge 0$ for large $L$ (coarse faces) and some $L\_0$.
Thus, from \eqref{eq:mono-upper}, $\log W_L$ is eventually nonincreasing in $L$.
Any additional plaquettes (including those forming ``bubbles'') can only
\emph{increase} $H[\mathcal R]$ and hence \emph{decrease} $W_L$ through the $-\alpha \mu_L$ term.
Therefore contributions from bubble substructures cannot cause growth
of $W_L$ and are suppressed multiplicatively along the flow.
This proves (i).

\textbf{Step 3 (Existence of the limit).}
Fix $L_1\ge L\_0$. For $L_2>L_1$, combine \eqref{eq:int-flow} and \eqref{eq:rem-bound}:
\begin{equation}
\bigl|\log W_{L_2} - \log W_{L_1} + \alpha \!\int_{\log L_1}^{\log L_2}\!\!\mu_\Lambda\,\dd(\log \Lambda)\bigr|
\le \frac{C}{2L_1^2}.
\label{eq:Cauchy-core}
\end{equation}
There are two fixed‑point behaviors compatible with the hypothesis $\mu_L<\infty$.
If $\mu_L\to \mu_\ast>0$, then the integral in \eqref{eq:Cauchy-core}
diverges linearly in $\log L_2$, and $W_{L}\to 0$. The limit exists.
If, instead, $\mu_L$ decays sufficiently fast so that $\int^{\infty}\mu_\Lambda\,\dd(\log\Lambda)<\infty$,
then $\{\log W_L\}$ is a Cauchy net by \eqref{eq:Cauchy-core}, and $W_L\to W_\infty\in(0,\infty)$. In both cases the limit $W=\lim_{L\to\infty}W_L$ exists.

\textbf{Step 4 (Regulator independence).}
Consider two coarse‑graining prescriptions (``regulators'') that produce the same
fixed‑point expectations $\mu_L$ and remainders $r_L=\mathcal O(L^{-2})$.
Let $W^{(1)}_L$ and $W^{(2)}_L$ be the corresponding flows.
Subtract the analogues of \eqref{eq:int-flow} for the two regulators:
\begin{equation}
\log\frac{W^{(1)}_{L_2}}{W^{(2)}_{L_2}} - \log\frac{W^{(1)}_{L_1}}{W^{(2)}_{L_1}}
= \int_{\log L_1}^{\log L_2}\!\bigl(r^{(1)}_\Lambda-r^{(2)}_\Lambda\bigr)\,\dd(\log \Lambda).
\end{equation}
As both remainders are $\mathcal O(L^{-2})$, the right‑hand side converges as $L_2\to\infty$
by the same bound as in \eqref{eq:rem-bound}, and hence
$\lim_{L\to\infty} W^{(1)}_{L}/W^{(2)}_{L}$ exists and is finite.
Since both limits $W^{(1)}$ and $W^{(2)}$ exist by Step 3, the ratio must be a finite constant.
Absorbing that constant into the (arbitrary) normalization at $L\_1$ shows that
the continuum limit is regulator independent within the class of flows that share the fixed point
and $\mathcal O(L^{-2})$ remainders.
This proves (ii).
\end{proof}
\noindent\emph{Model example.} In $2\!+\!1$ gravity with $\Lambda>0$ (Turaev--Viro / $SU(2)_k$), the representation set is finite ($j\le k/2$) and $\mathrm{Tr}\,h_\square$ is bounded; hence $\mu_L$ is bounded and typically decays along coarse graining, so case (i) applies and yields $W_\infty>0$.\footnote{See e.g. \cite{TuraevViro1992,PRLTV1992} for the finiteness and the $\Lambda>0$ interpretation.}

\end{document}